\documentclass[a4paper,12pt]{article} 
 \pdfoutput=1 
\usepackage{amsmath}
\usepackage{bm,color,url}
\usepackage{amssymb}
\usepackage{graphicx}
\usepackage{slashed}
\usepackage{ulem}
\usepackage[top=30truemm,bottom=30truemm,left=25truemm,right=25truemm]{geometry}

\begin{document}
\titlepage

\today

  \begin{flushright}
   {\bf
  \begin{tabular}{l}
OCHA-PP-350
  \end{tabular}
   }
  \end{flushright}

\vspace*{1.0cm}
\baselineskip 18pt
\begin{center}
 {\Large \bf
 Lepton flavor violation via four-Fermi contact interactions at the
 International Linear Collider 
} 

\vspace*{1.0cm} 

{\bf 
 Gi-Chol Cho$^a$, Yuka Fukuda$^b$, Takanori Kono$^a$
}

\vspace*{0.5cm}
$^a${\sl Department of Physics, Ochanomizu University, Tokyo 112-8610, Japan} \\
 $^b${\sl Graduate school of Humanities and Sciences, Ochanomizu University, Tokyo 112-8610, Japan} \\
\end{center}

\vspace*{1cm}

\baselineskip 18pt
\begin{abstract}
 \noindent

Lepton flavor violating (LFV) process $e^+ e^- \to e^+ \tau^-$ 
induced by the four-Fermi contact interactions at the International
 Linear Collider (ILC) is studied.
 Taking account of the event selection conditions, it is shown that the 
 ILC is sensitive to smaller LFV couplings as compared to
the measurement of $\tau\to 3e$ process at the B-factory experiment. 
 The upper bounds on some of the LFV couplings are improved by several
 factors using polarized $e^-/e^+$ beams at $\sqrt{s}=250~\mathrm{GeV}$ 
 and by an order of magnitude at $\sqrt{s}=1~\mathrm{TeV}$. 
\end{abstract}

\newpage
\section{Introduction}
Observation of neutrino oscillation~\cite{Fukuda:1998mi} tells us that
neutrinos have tiny mass and the lepton flavor is no longer
conserved. However, if neutrino mass is the only source of lepton flavor
violation (LFV), we may not be able to observe any LFV processes
since the size of phenomena is too small to measure experimentally. 
This is because the lepton flavor symmetry is restored in the limit of
massless neutrino.
For example, the branching fraction of LFV decay process $\mu \to
e\gamma$ with non-zero neutrino mass is estimated as
\begin{align}
 \mathrm{Br}(\mu\to e\gamma)=\frac{3}{32\pi}\alpha
 \left| U_{ei}^* U_{\mu i} \right|^2
 \left(\frac{m_{\nu_i}}{m_W}\right)^4
 < 10^{-48}
 \left(\frac{m_{\nu_i}}{1~\mathrm{eV}} \right)^4,
 \label{megam}
\end{align}
where $\alpha,U_{ij}$ and $i$ denote the fine-structure constant,
lepton-flavor mixing matrix and the generation index, respectively.

On the other hand, some new physics models may have sources of LFV
beyond the neutrino mass, and it might be expected to observe sizable
LFV effects at future experiments.
No signature of new physics, however, has been found yet at the LHC
experiments, so an interpretation of data at the LHC suggests that a
scale of new physics is much higher than $\mathcal{O}$(TeV). 
In such a case,
it is known that mediation by a heavy particle at collider processes 
are described by four-Fermi contact interactions in a good
approximation. 

We study LFV process $e^+ e^- \to e^+ \tau^-$ via
four-Fermi contact interactions at the International Linear 
Collider (ILC)~\cite{Baer:2013cma}. 
One of the authors (G.C.C.) has studied possibilities to search for LFV
contact interactions in $e^+ e^- \to e^+ \ell^-$ and $e^-
e^- \to e^- \ell^-$ at the ILC~\cite{Cho:2016zqo} where $\ell$ stands
for $\mu$ or $\tau$. It was pointed out that present bounds on LFV
contact interactions for $\ell=\mu$ from the measurement of
$\mathrm{Br}(\mu \to 3e)$ at SINDRUM experiment~\cite{Bellgardt:1987du}
are stringent so that no improvement of constraints on the LFV contact
interactions is expected at the ILC.
Moreover, bounds on LFV interactions for $\ell=\tau$ were obtained
merely by naive evaluation of signal and background cross sections at
parton level. 

In this paper, we introduce a selection using the variable $m_T^{\ell}$
in order to reduce the background process efficiently and investigate 
the sensitivity of ILC for the LFV contact interaction taking account of
detector effects. 
Moreover, we show that appropriate use of lepton beam polarizations
allows to reduce SM background and thus increases the sensitivity to 
some of the LFV parameters, 
although the polarization may have negative effect on signal events
for certain parameters. 
We show that, even at the first stage of 
the ILC experiment, i.e., $\sqrt{s}=250~\mathrm{GeV}$ with the 
integrated luminosity $2~\mathrm{ab}^{-1}$~\cite{Fujii:2017vwa},
the upper bounds on some LFV contact 
interactions could be improved by several factors compared to the 
previous bounds from $\mathrm{Br}(\tau \to 3e)$ at the Belle 
experiment~\cite{Hayasaka:2010np}.
The improvement of the bounds could be nearly an order of magnitude for
$\sqrt{s}=1~\mathrm{TeV}$. 

There are some previous studies on this topic, e.g.,  
refs.~\cite{Ferreira:2006dg,Murakami:2014tna}.
Those works have been done in the parton level without taking account of
the hadronization/reconstruction efficiency of the $\tau$-lepton at the
detector. 
The effect of beam polarization has not been evaluated and 
the analysis has been performed on only vector-type contact 
interactions~\cite{Murakami:2014tna} while scalar-type interactions are
examined in our study. 

This paper is organized as follows. In Sec.~II, we briefly review our
effective Lagrangian given by LFV contact interactions and some
observables at experiments. Results of numerical analysis and
constraints on LFV contact interactions will be given in Sec.~III.
We give some discussions in Sec.~IV. 
Sec.~V is devoted to summarize our work.

 \section{Effective Lagrangian} 

 The effective Lagrangian which describes the LFV process $e^+ e^- \to
 e^+ \tau^-$ via contact interactions 
 consists of the following six operators after the Fierz
 rearrangement~\cite{Kuno:1999jp}
\begin{align}
 {\mathcal{L}_{\mathrm{eff}}}&=
 -\frac{4G_F}{\sqrt{2}}
 \left\{
  g_1 \left(\overline{\tau_R}e_L \right) \left(\overline{e_R}e_L\right)
   +
  g_2 \left(\overline{\tau_L}e_R \right) \left(\overline{e_L}e_R \right)      
  \right.
\nonumber \\
 &+
  g_3 \left(\overline{\tau_R}\gamma^\mu e_R \right)
  \left(\overline{e_R}\gamma_\mu e_R \right)      
  +
  g_4 \left(\overline{\tau_L}\gamma^\mu e_L \right)
  \left(\overline{e_L}\gamma_\mu e_L \right)      
\nonumber \\
 &+
 \left.
  g_5 \left(\overline{\tau_R}\gamma^\mu e_R \right)
  \left(\overline{e_L}\gamma_\mu e_L \right)      
  +
  g_6 \left(\overline{\tau_L}\gamma^\mu e_L \right)
  \left(\overline{e_R}\gamma_\mu e_R \right)      
 \right\}+\mathrm{h.c.},
 \label{eq:efflag}
\end{align}
where $G_F$ denotes the Fermi coupling constant, and 
the subscripts $L$ and $R$ represent the chirality of a fermion $f$,
i.e., $f_{L(R)}\equiv \frac{1-(+)\gamma_5}{2}f$.
The six couplings $g_i~(i=1\sim 6)$ are dimensionless parameters. 
The first two terms in eq.~(\ref{eq:efflag}) are the scalar-type
interactions, while the others are the vector-type interactions.


In the limit of massless leptons, 
the spin-averaged differential cross section in the center-of-mass (CM) 
system for $e^+ e^- \to e^+ \tau^-$ is 
calculated from the effective Lagrangian~(\ref{eq:efflag}) as  
\begin{align}
 \frac{d \sigma(e^+ e^- \to e^+ \tau^-)}{d\cos\theta}
  &=
  \frac{G_F^2 s}{64\pi}
  \left[
   \left( G_{12}+16G_{34} \right) (1+\cos\theta)^2
  +4G_{56}\left\{4+(1-\cos\theta)^2 \right\}
     \right], 
\label{diffxepem}
\end{align}
where the parameter $G_{ij}$ is defined as
\begin{align}
 G_{ij}\equiv |g_i|^2+  |g_j|^2.  
\end{align}
Integrating (\ref{diffxepem}) over $\cos\theta$, the cross section is
given as
\begin{align}
 \sigma = \frac{G_F^2 s}{24\pi}
 \left\{
 G_{12} + 16 (G_{34}+G_{56})
 \right\}
 &\approx
 4.4~\mathrm{fb}
\left(
 \frac{\sqrt{s}}{250~\mathrm{GeV}}
 \right)^2
 \left\{
 \frac{ G_{12}+16(G_{34}+G_{56})}
{10^{-4}}
 \right\}
 \nonumber \\
 &= 
 4.4~\mathrm{fb}
\left(
 \frac{\sqrt{s}}{250~\mathrm{GeV}}
 \right)^2
\sum_{i=1}^6
 a_i
 \left( \frac{ g_i}{10^{-2}}\right)^2, 
\label{xsect} 
\end{align}
where a coefficient $a_i$ is given by
1 for $i=1,2$ and 16 for $i=3\sim 6$, respectively. 

The LFV process $\tau \to eee$ ($\tau \to 3e$) is also given by 
the same effective Lagrangian~(\ref{eq:efflag}).  
The branching ratio of $\tau \to 3e$ is expressed in terms of $G_{ij}$
as
\begin{align}
\mathrm{Br}(\tau\to 3e)
&=\frac{\tau_\tau}{\tau_\mu}
\left(\frac{m_\tau}{m_\mu}\right)^5
\times \frac{1}{8}
\left(G_{12}+16 G_{34}+8 G_{56}\right)
\nonumber \\
 &\approx 0.022 \times \left(G_{12}+16 G_{34}+8 G_{56}\right),
  \label{brtau3l}
\end{align}
where $\tau_\tau$ and $\tau_\mu$ are the lifetime of $\tau$ and $\mu$,
respectively, and we adopt $\tau_\tau=2.91 \times 10^{-13}~\mathrm{s}$
and $\tau_\mu=2.20 \times 10^{-6}~\mathrm{s}$ for the numerical
evaluation~\cite{Agashe:2014kda}.
The upper bound on $\mathrm{Br}(\tau \to 3 e)$ at 90\% CL 
has been given by the Belle collaboration as~\cite{Hayasaka:2010np} 
\begin{align}
 \mathrm{Br}(\tau^-\to e^- e^+ e^-)
 &< 2.7 
\times 10^{-8}, 
  \label{belle_eee}
\end{align}
and the bound (\ref{belle_eee}) can be read as upper limits on the LFV
couplings $G_{ij}$ as
  \begin{align}
   \left\{
   G_{12},\, G_{34},\,  G_{56}
   \right\}
  <
\left\{
  1.2\times 10^{-6}, \,
  7.5\times 10^{-8},\,
  1.5\times 10^{-7}
   \right\}. 
  \label{upp_belle}
  \end{align}
We give upper bounds (\ref{upp_belle}) in terms of the LFV coupling
$g_i$ for later convenience:
\begin{align}
 g_1, g_2 < 1.1\times 10^{-3}, \,\,
 g_3, g_4 < 2.7\times 10^{-4}, \,\,
 g_5, g_6 < 3.9\times 10^{-4}.
 \label{eq:bound1}
\end{align}
Throughout this paper, we adopt the single-coupling dominant hypothesis 
to examine the upper limit on the LFV couplings. For example, the
limits~(\ref{eq:bound1}) are obtained by allowing only one coupling
among six $g_i$ is finite while other five couplings are set to zero. 
The size of the cross section (\ref{xsect}) at
$\sqrt{s}=250~\mathrm{GeV}$ corresponding to the limits obtained by the
Belle collaboration (\ref{upp_belle}) can be estimated as 
  $5.3\times 10^{-2}~\mathrm{fb}, 4.8\times 10^{-2}~\mathrm{fb}$ and
  $1.1\times 10^{-1}~\mathrm{fb}$, respectively. 
 The limits of branching fraction (\ref{belle_eee}) and other various 
 LFV decay modes of $\tau$-lepton are expected to improve by one or two
  orders of magnitude at the super-KEKB~\cite{Bevan:2014iga}.
 \section{Constraints on LFV couplings at the ILC}
 In this section, we investigate sensitivity on the LFV couplings in
 (\ref{eq:efflag}) at the ILC. Throughout our analysis, we use 
 {\sc MadGraph5\_aMC@NLO}~\cite{Alwall:2014hca}, 
 {\sc PYTHIA~8}~\cite{Sjostrand:2007gs} and  
 {\sc DELPHES~3}~\cite{deFavereau:2013fsa} 
 for event generations, hadronization and detector simulation,
 respectively. 
 The limits on the LFV couplings are estimated using {\sc
 MadAnalysis~5}~\cite{Conte:2012fm}. 

In the detector simulation, we assume a detector like the ILD
(International Large Detector) at the ILC, which is expected to have
good performance on the reconstruction of the $\tau$-lepton. 
The reconstruction efficiency in leptonic decay modes could be about 
99\%~\cite{Behnke:2013lya} in the pseudorapidity ($\eta$) range $|\eta|<2.4$. 
In the hadronic decay modes such as
$\tau \to \pi \nu,~\rho(\to 2\pi) \nu,~a_1(\to 3\pi) \nu$,
the reconstruction efficiency is estimated as 95\% for $\pi$ and 90\%
for $\rho$ and $a_1$ modes, respectively~\cite{Tran:2015nxa}.

\begin{figure}[t]
  \begin{center}
   \includegraphics[clip, width=9cm]{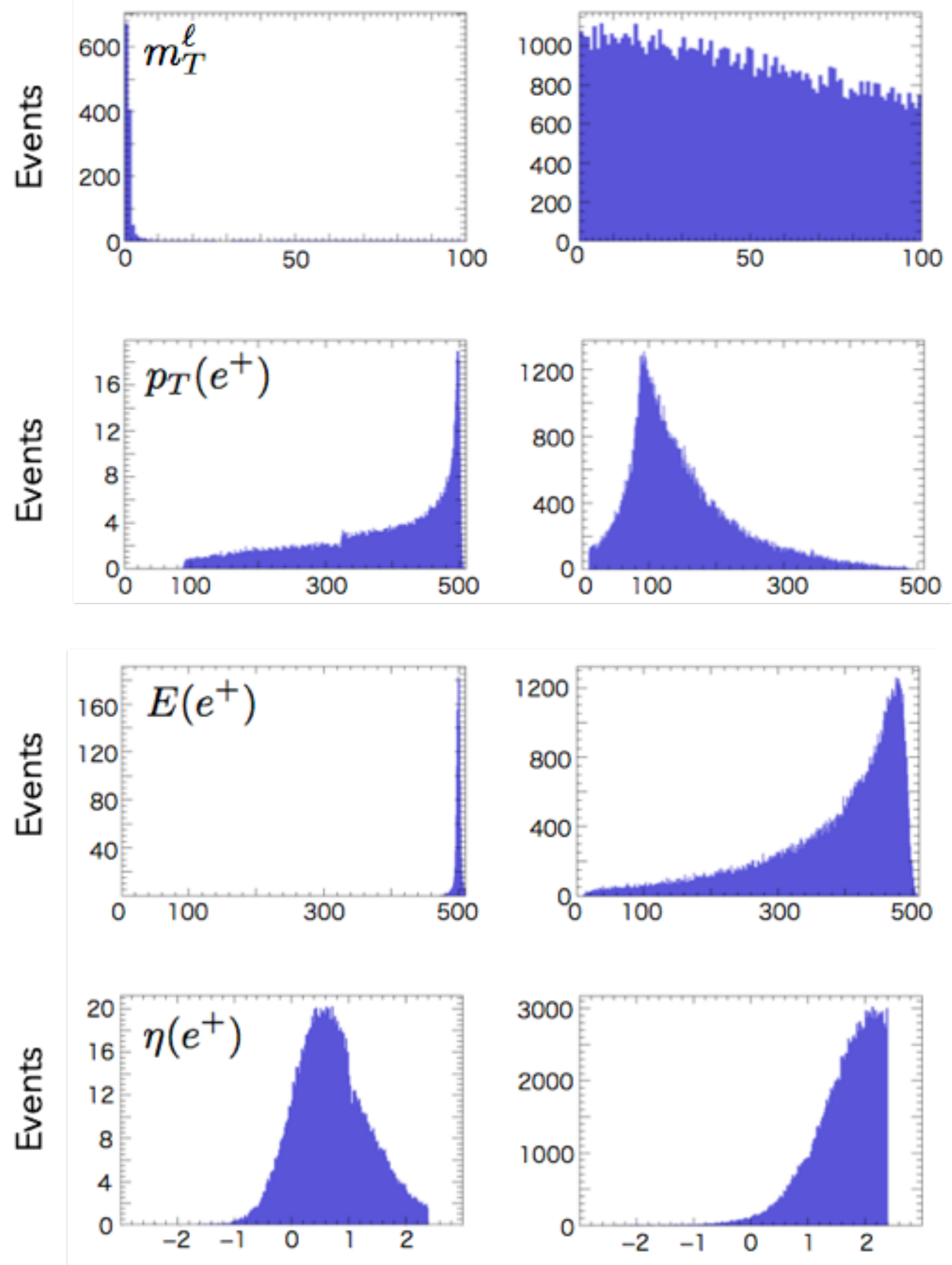}
   \caption{
   Event distributions of the transverse mass $m_T^\ell$, and 
   the energy $E$, the transverse momentum $p_T$
   and the pseudo rapidity $\eta$ for $e^+$.
   The unit of horizontal axis in $m_T^\ell$, 
   $E(e^+)$ and $p_T(e^+)$ is GeV. Plots in the left- and the
   right-sides are signal and background processes, respectively.
   Both events are evaluated for $\sqrt{s}=1~\mathrm{TeV}$ and
   $L_\mathrm{int}=2~\mathrm{ab}^{-1}$. 
   The LFV couplings in the signal events are set for 
   $g_1=10^{-3}$ and $g_2\sim g_6=0$. 
   }
    \label{fig:sig_bg}
  \end{center}
\end{figure}

As a background to the signal process $e^{+}e^{-}\to e^{+}\tau^{-}$, we consider
the SM process $e^+ e^- \to e^+ \nu_e \tau^- \bar{\nu}_{\tau}$.
Among the diagrams generating this final state, the largest contribution comes
from the diagram where the $W^{-}$ boson, which is radiated off the $e^{-}$ in
$e^{+}e^{-}\to e^{+}e^{-}$, transforms into $\tau^{-} \bar{\nu}_{\tau}$.
We apply several event selection conditions to reduce contribution from this background process.
As an example, we compare several event 
distributions of the signal and background processes at
$\sqrt{s}=1~\mathrm{TeV}$ with an integrated luminosity of
$L_\mathrm{int}=2~\mathrm{ab}^{-1}$ in Fig.~\ref{fig:sig_bg}. 
Signal events in the figure are obtained for the LFV 
couplings $g_1=10^{-3}$ and $g_i=0~(i=2\sim 6)$. 
The first row in Fig.~\ref{fig:sig_bg} shows
the distributions of the transverse mass $m_T^\ell$ defined as 
\begin{align}
 ( m_T^{\ell})^2 =
 ( |\bm{p}_T^{\ell^-}| + |\slashed{\bm{p}}_T|)^2
-
 (\bm{p}_T^{\ell^-} + \slashed{\bm{p}}_T)^2, 
  \label{trsvsm}
\end{align}
where $\bm{p}_T^{\ell^-}$ and $\slashed{\bm{p}}_T$ represent the 
transverse momentum of $\ell^-(=e^-,\,\mu^-)$ from the $\tau^-$ decay 
and the missing transverse momentum, respectively. 
Since the origin of $\slashed{\bm{p}}_T$ in the signal event is
neutrinos ($\nu_\tau,~\overline{\nu}_{\ell}$) from the $\tau$ decay,
the $m_T^\ell$ distribution has a peak below $m_\tau$ while the 
background process does not have such a peak due to the extra
$\nu_e$ in addition to ($\nu_\tau,~\overline{\nu}_{\ell}$) from the
$\tau$ decay.
In the figure, we also compare 
the transverse momentum, energy and pseudorapidity of $e^+$ in the
final state, respectively. 
Taking account of differences between signal and background distributions
in Fig.~\ref{fig:sig_bg},
we require the following conditions on $m_T^\ell$ and $p_T(e^+)$: 
 \begin{align}
  m_T^\ell &\le 10~\mathrm{GeV},
\\
  p_T(e^+) &\ge 70,~150,~300~\mathrm{GeV}~
  (\sqrt{s}=250~,500~\mathrm{GeV},~1~\mathrm{TeV}). 
 \end{align}
As an example, we show the number of events in signal and background 
(i) before applying any cuts, (ii) requiring $m_T^\ell \le 10~\mathrm{GeV}$,
(iii) further requiring $p_T(e^+)\ge 300~\mathrm{GeV}$, in case of 
$g_1=10^{-3}$, $\sqrt{s}=1~\mathrm{TeV}$ and 
$L_{\mathrm{int}}=2~\mathrm{ab}^{-1}$ in Table~\ref{tab_ex}. 
With these selections, the contribution of the background process is reduced
by a factor of 200 while keeping almost 60\% of signal events.
  \begin{table}[h]
\begin{center}
 \begin{tabular}{c|c|c}
  \hline \hline
  & signal & background
      \\ 
  \hline
  (i)  & 1368 & 171120
      \\
 (ii) & 1167 & 9976
      \\
  (iii) & 779 & 801
      \\
\hline \hline
 \end{tabular}
 \caption{
 Number of events in signal and background events for
 $g_1=10^{-3},~\sqrt{s}=1~\mathrm{TeV}$ and
 $L_{\mathrm{int}}=2~\mathrm{ab}^{-1}$.
 Conditions (i)$\sim$(iii) are:
 (i) no cut, (ii) $m_T^\ell \le 10~\mathrm{GeV}$,
 (iii) $p_T(e^+)\ge 300~\mathrm{GeV}$ and
 $m_T^\ell \le 10~\mathrm{GeV}$. 
}
 \label{tab_ex}
 \end{center}
\end{table}


We define the statistical significance by $S/\sqrt{B}$, where $S$ and
$B$ denote the number of signal and background events, respectively.
We set the upper bounds on the LFV couplings $g_i$ by requiring
$S/\sqrt{B}=2$, which correspond to the upper limits at the 95.4\%
confidence level.
The number of events in both signal and background processes is
obtained for an integrated luminosity of 
$L_\mathrm{int}=2~\mathrm{ab}^{-1}$~\cite{Fujii:2017vwa}. 
\begin{figure}[t]
  \begin{center}
   \includegraphics[clip, width=16cm]{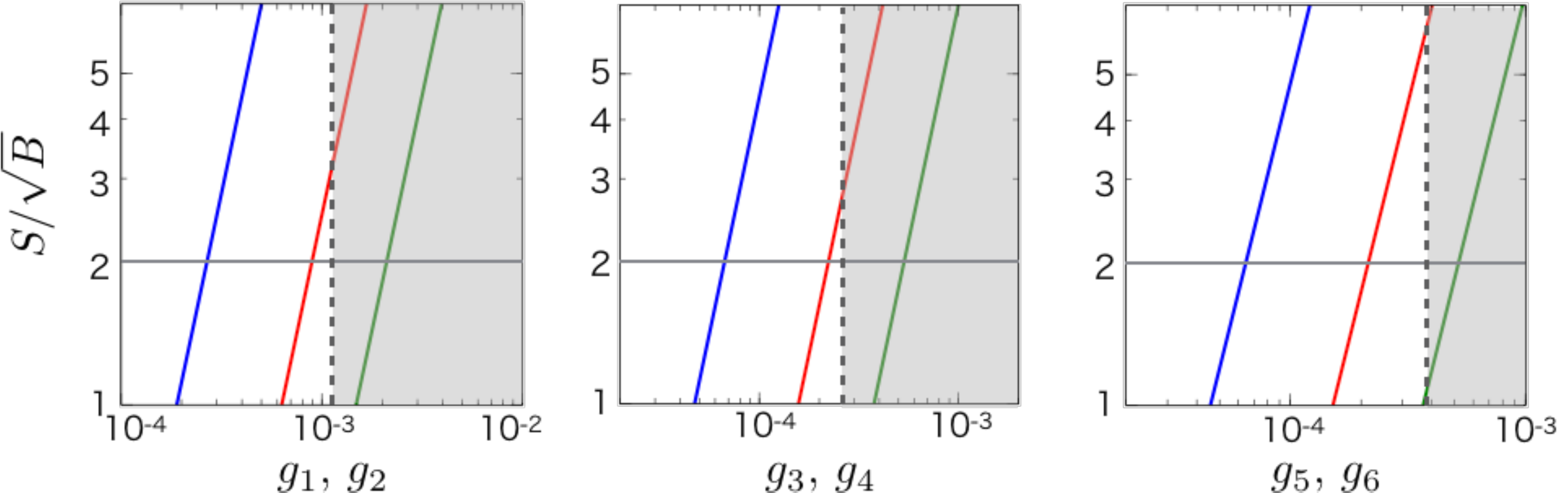}
   \caption{
   Statistical significance $S/\sqrt{B}$ as function of the LFV couplings.
   Three colored lines correspond to $\sqrt{s}=250~\mathrm{GeV}$
   (green), $500~\mathrm{GeV}$ (red) and $1~\mathrm{TeV}$ (blue),
   respectively. The integrated luminosity is fixed at
   $L_\mathrm{int}=2~\mathrm{ab}^{-1}$. The shaded region stands for the
   excluded range of $g_i$ from $\mathrm{Br}(\tau \to 3e)$ at
   Belle~\cite{Hayasaka:2010np}. 
   }
    \label{fig:sig_bg_cut}
  \end{center}
\end{figure}
In Fig.~\ref{fig:sig_bg_cut},
we show $S/\sqrt{B}$ as a function of the LFV couplings $g_i$.
Since, as shown in eq.~(\ref{diffxepem}), there are three pairs of six
LFV couplings as independent parameters in the differential cross
section, we show three plots for $(g_1,g_2)$, $(g_3,g_4)$ and
$(g_5,g_6)$ separately. 
Three colored lines stand for $\sqrt{s}=250~\mathrm{GeV}$ (green),
$500~\mathrm{GeV}$ (red) and $1~\mathrm{TeV}$ (blue), respectively.
Shaded region in three plots represent excluded range of the LFV
coupling $g_i$ from the measurement of $\mathrm{Br}(\tau \to 3e)$ by the
Belle collaboration~\cite{Hayasaka:2010np}. 
Since the cross section of the signal process is 
proportional to the center-of-mass (CM) energy as shown in
eq.~(\ref{xsect}), the upper bounds on the LFV couplings are much severe
for larger $\sqrt{s}$.
We find that better bounds can be obtained at the ILC 
with $\sqrt{s}=500~\mathrm{GeV}$ and $1~\mathrm{TeV}$ for all
six parameters, compared to those from the Belle experiment. 
On the other hand, for $\sqrt{s}=250~\mathrm{GeV}$, the upper limits on 
only $g_5,~g_6$ could be better (but marginal).
These limits on $g_i$ are obtained with the integrated luminosity of 
$L_\mathrm{int}=2~\mathrm{ab}^{-1}$. 
For different $L_\mathrm{int}$, the limits on the LFV couplings are
scaled by a factor $\left( 2~\mathrm{ab}^{-1}/L_\mathrm{int}\right)^{1/4}$ 
since the significance $S/\sqrt{B}$ is proportional to 
$\sqrt{L_\mathrm{int}}$ while the number of signal events $S$ is 
proportional to the LFV coupling $g_i^2$.

\begin{figure}[t]
  \begin{center}
   \includegraphics[clip, width=16cm]{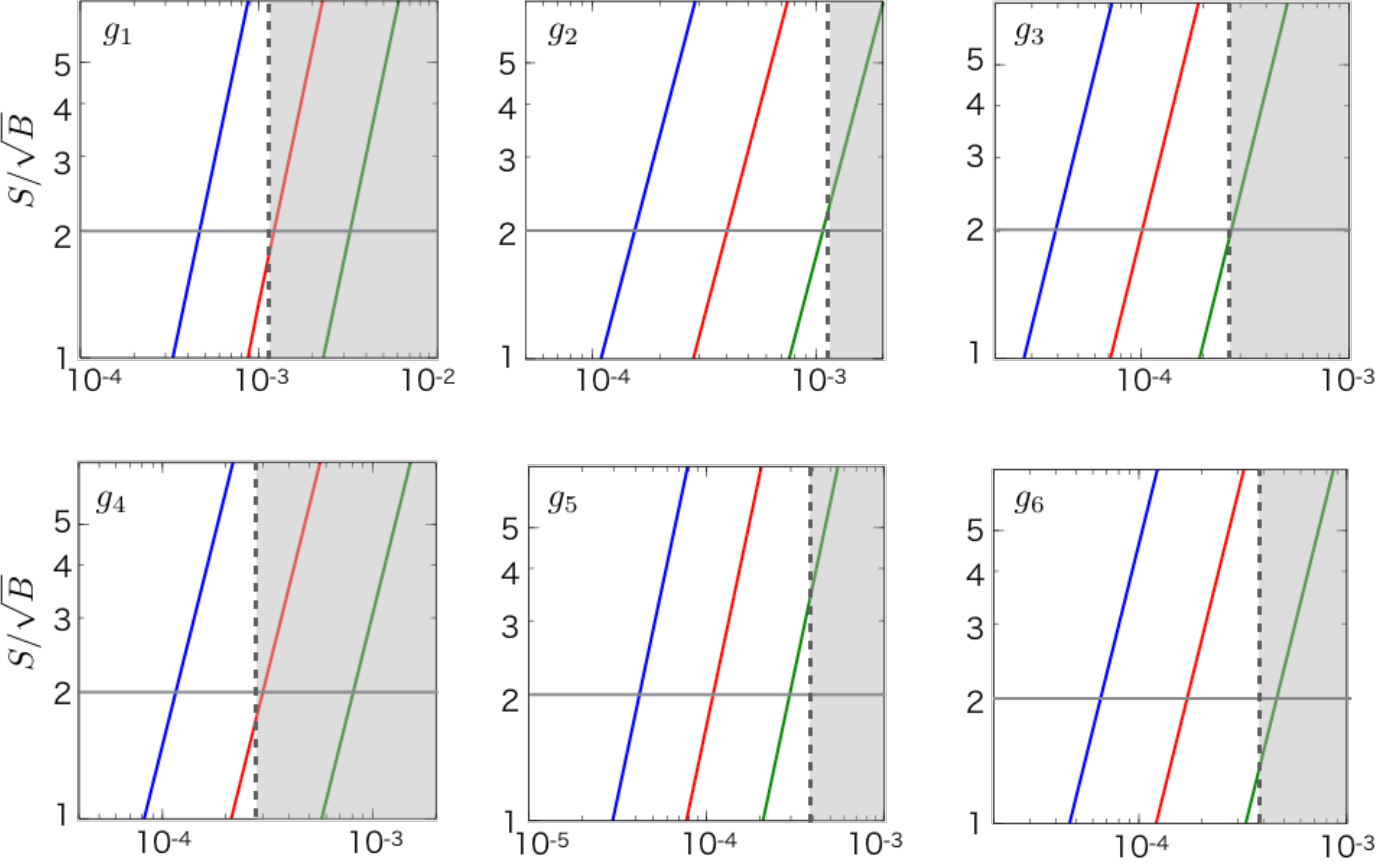}
   \caption{
   $S/\sqrt{B}$ as a function of the LFV coupling $g_i$ with the
   $e^-$ beam polarization $P(e^-)=0.8$.
   The description on the colored lines and 
   shaded region are the same with those in Fig.~\ref{fig:sig_bg_cut}.
   }
    \label{fig:sig_bg_pol}
  \end{center}
\end{figure}

\begin{figure}[t]
  \begin{center}
   \includegraphics[clip, width=16cm]{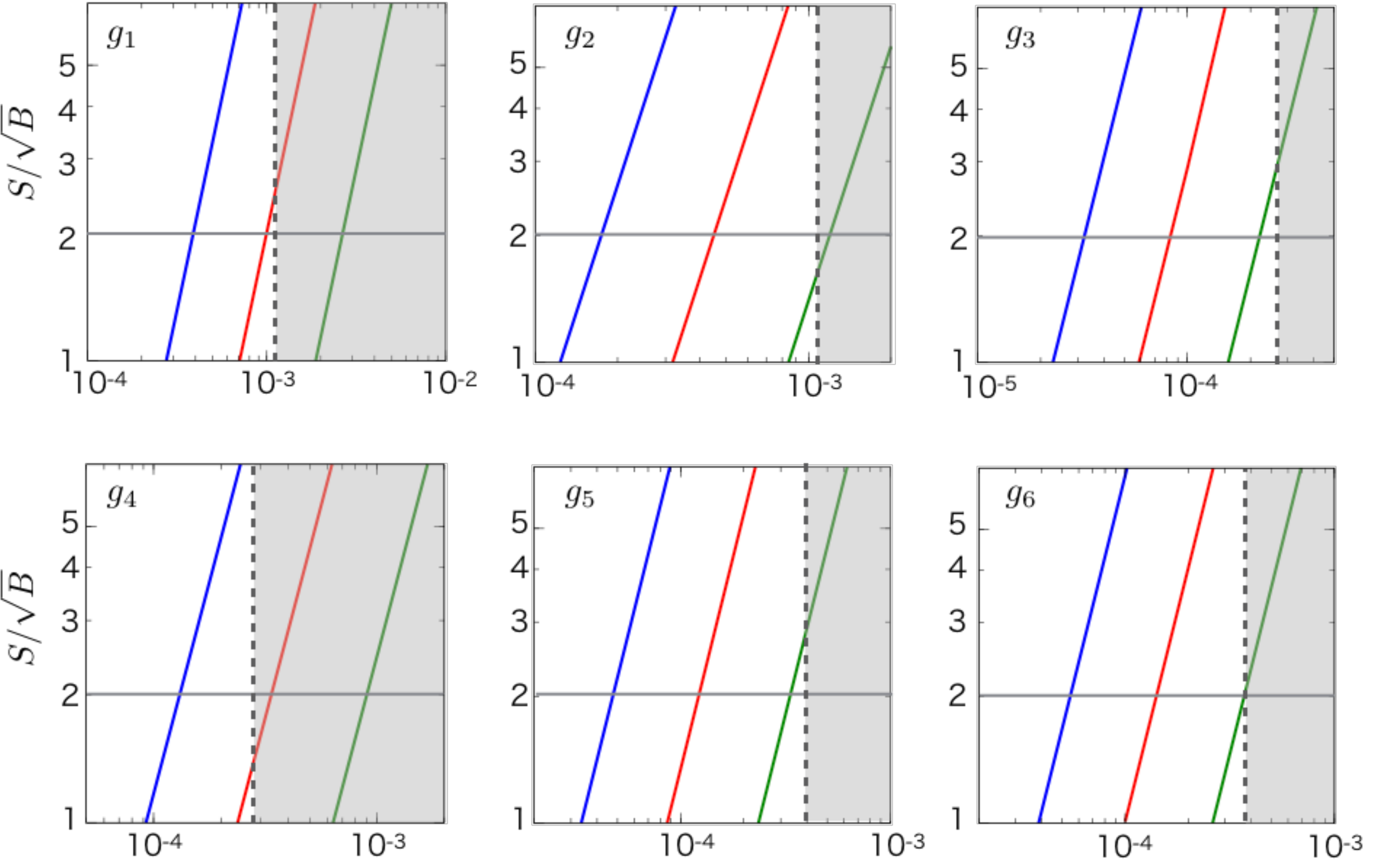}
   \caption{
   $S/\sqrt{B}$ as a function of the LFV coupling $g_i$ with the
   beam polarization $P(e^-)=0.8$ and $P(e^+)=-0.3$.
   The description on the colored lines and 
   shaded region are the same with those in Fig.~\ref{fig:sig_bg_cut}.
   }
    \label{fig:sig_bg_polpol}
  \end{center}
\end{figure}

Next we discuss the possibility to use polarized initial beams at the ILC.
In some of the SM background processes, the initial $e^-$ couples to the
$W$-boson. Such processes could be suppressed efficiently by polarizing
the $e^-$ beam to be right-handed since $W$-boson couples only to
left-handed fermions.
As the SM background process is dominated
by the initial helicity state $e^-_L+e^+_R$
(97\% at $\sqrt{s}=250~\mathrm{GeV}$ and 
92\% at $\sqrt{s}=500~\mathrm{GeV}$ or $1~\mathrm{TeV}$), the ILC has
a stronger sensitivity to couplings with $e^-_R$ and $e^+_L$ where SM 
contribution is highly suppressed. 
In Fig.~\ref{fig:sig_bg_pol} we give the significance $S/\sqrt{B}$ for
each LFV couplings with polarized electron beam. We set the polarization
of the $e^-$ beam to 80\% (hereafter we denote it as $P(e^-)=0.8$).  
With this condition, the cross section of the background process is
reduced to 17~fb from 86~fb in the unpolarized case. 
The upper bounds on some of LFV couplings at $S/\sqrt{B}=2$  
are much improved as compared to the unpolarized case
(Fig.~\ref{fig:sig_bg_cut}) due to the
suppression of the background processes mediated by the $W$-boson as we
expected. 
For example, the upper limits on $g_2,\,g_3$ and $g_5$ are better
than those from the Belle experiment even for 
$\sqrt{s}=250~\mathrm{GeV}$. Those improvements could be an order of
magnitude for $\sqrt{s}=1~\mathrm{TeV}$.
On the other hand,
the upper limits on the LFV couplings $g_1$ and $g_4$ are worse than the
unpolarized case.
Since the chirality of initial electron is left-handed
in the operators with the coupling $g_1$ or $g_4$,
these operators do not contribute to the LFV processes when
the electron is polarized to be right-handed, thus causing suppression 
of the event rate.
We also show the results where both $e^-$ and $e^+$ beams are polarized
as $P(e^-)=0.8,~P(e^+)=-0.3$ in Fig.~\ref{fig:sig_bg_polpol}.
The limits on $g_1\,g_3$ and $g_6$ are marginally better than the
previous case ($P(e^-)=0.8$) while $g_2,\,g_4$ and $g_5$ are worse.  
The upper limits of the LFV couplings at 95\% CL in all cases
(unpolarized, polarized beams) are summarized in
Table~\ref{table:limits}.
\begin{table}[h]
\begin{center}
  \begin{tabular}{@{\vrule width 1pt}r|c|c|c|c|c|c@{\ \vrule width 1pt}}
\hline \hline
  & $g_1$ & $g_2$ & $g_3$ & $g_4$ & $g_5$ & $g_6$
         \\[0.2mm]
   \hline
   Belle~\cite{Hayasaka:2010np}
   & \multicolumn{2}{c|}{$1.1 \times 10^{-3}$}
   & \multicolumn{2}{c|}{$2.7 \times 10^{-4}$}
   & \multicolumn{2}{c@{\ \vrule width 1pt}}{$3.9 \times 10^{-4}$}
          \\[0.2mm]   \hline
   Belle II~\cite{Kou:2018nap}
   & \multicolumn{2}{c|}{$1.3 \times 10^{-4}$}
   & \multicolumn{2}{c|}{$3.2 \times 10^{-5}$}
   & \multicolumn{2}{c@{\ \vrule width 1pt}}{$4.7 \times 10^{-5}$}
           \\[0.2mm]
   \hline    \hline
   & \multicolumn{6}{c@{\ \vrule width 1pt}}{unpolarized}
         \\[0.2mm]
   \hline
   $\sqrt{s}=250~\mathrm{GeV}$
   & \multicolumn{2}{c|}{$2.2 \times 10^{-3}$}
   & \multicolumn{2}{c|}{$5.4 \times 10^{-4}$}
   & \multicolumn{2}{c@{\ \vrule width 1pt}}{$5.3 \times 10^{-4}$}
           \\[0.2mm]
   \hline
   $500~\mathrm{GeV}$
   & \multicolumn{2}{c|}{$9.1 \times 10^{-4}$}
   & \multicolumn{2}{c|}{$2.2 \times 10^{-4}$}
   & \multicolumn{2}{c@{\ \vrule width 1pt}}{$2.2 \times 10^{-4}$}
           \\[0.2mm]
   \hline
   $1~\mathrm{TeV}$
   & \multicolumn{2}{c|}{$2.8 \times 10^{-4}$}
   & \multicolumn{2}{c|}{$6.8 \times 10^{-5}$}
   & \multicolumn{2}{c@{\ \vrule width 1pt}}{$6.6 \times 10^{-5}$}
					   \\[0.2mm]
   $\star~\cite{Cho:2016zqo}$
   & \multicolumn{2}{c|}{$9.4 \times 10^{-4}$}
   & \multicolumn{2}{c|}{$2.4 \times 10^{-4}$}
   & \multicolumn{2}{c@{\ \vrule width 1pt}}{$2.3 \times 10^{-4}$}
              \\[0.2mm]
   \hline
   & \multicolumn{6}{c@{\ \vrule width 1pt}}{polarized ($P(e^-)=0.8$)}
         \\[0.2mm]
   \hline
   $\sqrt{s}=250~\mathrm{GeV}$
   &$3.3\times 10^{-3}$ &$1.1\times 10^{-3}$ 
   &$2.7\times 10^{-4}$ &$8.2\times 10^{-4}$
   &$3.0\times 10^{-4}$ &$4.7\times 10^{-4}$           
           \\[0.2mm]
   \hline
   $500~\mathrm{GeV}$
   &$1.2\times 10^{-3}$ &$4.0\times 10^{-4}$ 
   &$1.0\times 10^{-4}$ &$3.0\times 10^{-4}$
   &$1.1\times 10^{-4}$ &$1.7\times 10^{-4}$
         \\[0.2mm]
   \hline
   $1~\mathrm{TeV}$
   &$4.8\times 10^{-4}$ &$1.6\times 10^{-4}$ 
   &$3.9\times 10^{-5}$ &$1.2\times 10^{-4}$
   &$4.3\times 10^{-5}$ &$6.6\times 10^{-5}$
         \\[0.2mm]
   \hline
   & \multicolumn{6}{c@{\ \vrule width 1pt}}{polarized
     ($P(e^-)=0.8$, $P(e^+)=-0.3$)}
         \\[0.2mm]
   \hline
   $\sqrt{s}=250~\mathrm{GeV}$
   &$2.7\times 10^{-3}$ &$1.2\times 10^{-3}$ 
   &$2.2\times 10^{-4}$ &$9.1\times 10^{-4}$
   &$3.4\times 10^{-4}$ &$3.8\times 10^{-4}$           
           \\[0.2mm]
   \hline
   $500~\mathrm{GeV}$
   &$1.0\times 10^{-3}$ &$4.6\times 10^{-4}$ 
   &$8.4\times 10^{-5}$ &$3.4\times 10^{-4}$
   &$1.2\times 10^{-4}$ &$1.4\times 10^{-4}$
         \\[0.2mm]
   \hline
   $1~\mathrm{TeV}$
   &$4.0\times 10^{-4}$ &$1.8\times 10^{-4}$ 
   &$3.3\times 10^{-5}$ &$1.3\times 10^{-4}$
   &$4.8\times 10^{-5}$ &$5.5\times 10^{-5}$
         \\[0.2mm]
\hline \hline
 \end{tabular}
 \caption{
 Summary of the upper limits on the LFV couplings at 95\% CL. 
 The limits for $\sqrt{s}=1~\mathrm{TeV}$ in ref.~\cite{Cho:2016zqo} 
 are shown for comparison in the row with the symbol $\star$. 
}
 \label{table:limits}
\end{center}
\end{table}

\section{Discussions}
\begin{figure}[htpb]
   \includegraphics[clip, width=16cm]{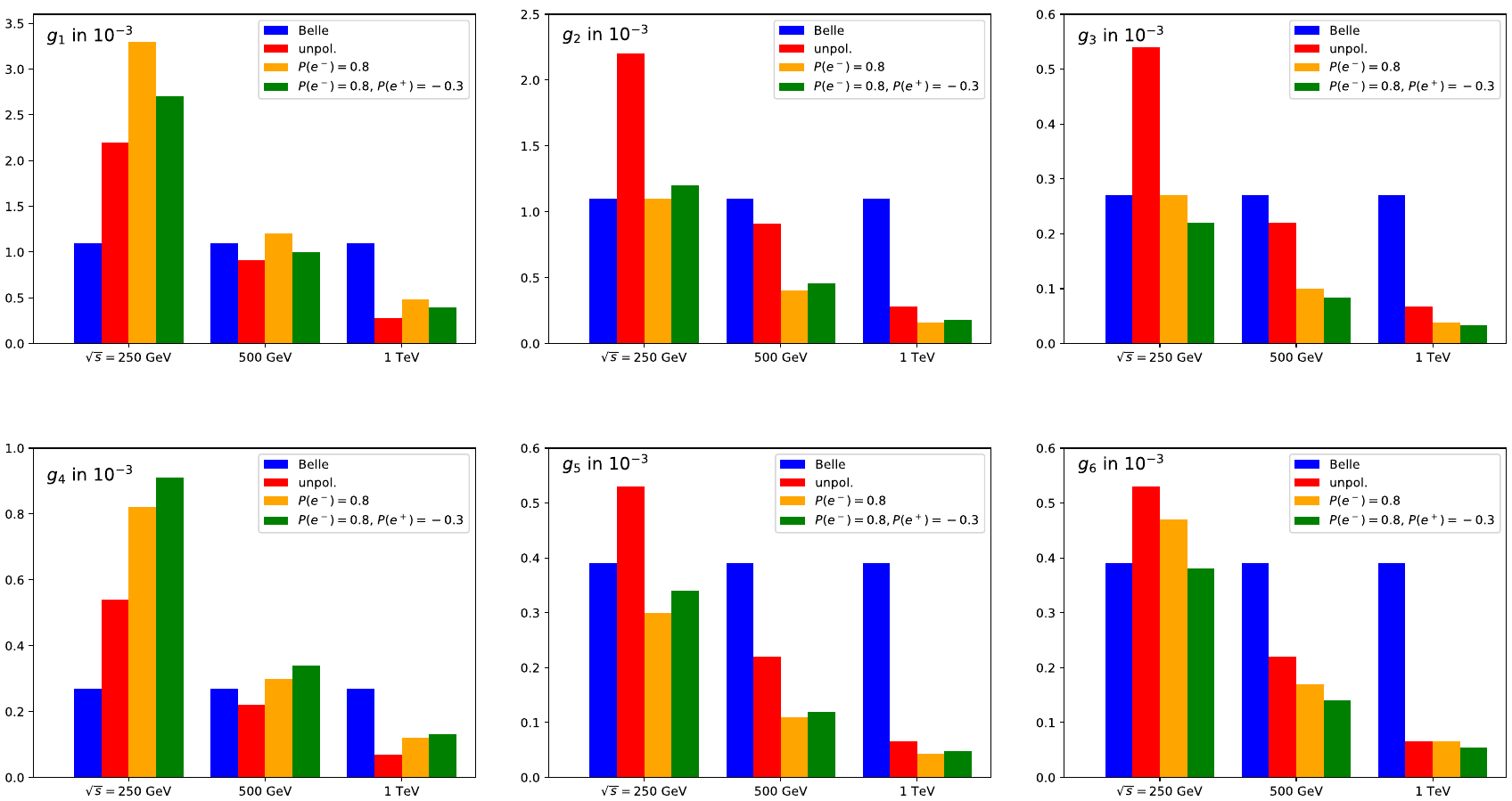}
 \caption{Summary of limits on the LFV couplings $g_i$ for
 $\sqrt{s}=250~\mathrm{GeV},~500~\mathrm{GeV}$ and $1~\mathrm{TeV}$.
 The bounds from the Belle experiment are shown in blue for comparison.
 The limits on $g_i$ are shown for three cases: unpolarized beam (red),
 polarized $e^-$ beam with $P(e^-)=0.8$ (orange), polarized both $e^-$
 and $e^+$ beams with $(P(e^-), P(e^+))=(0.8, -0.3)$ (green). 
 }
    \label{fig:sig_bg_bar} 
 \end{figure}

Results of our study are summarized in Fig.~\ref{fig:sig_bg_bar}.
In the figure, we compare the upper limits on the couplings from 
$\tau \to 3e$ with three cases in the ILC experiment;
(i) unpolarized beam, (ii) polarized $e^-$ beam with $P(e^-)=0.8$,
 and (iii) polarized $e^-$ and $e^+$ beams with $P(e^-)=0.8$ and
 $P(e^+)=-0.3$.
 The CM energy dependence 
 ($\sqrt{s}=250~\mathrm{GeV},~500~\mathrm{GeV}$ and $1~\mathrm{TeV}$) 
  of the upper limits is also shown in each plot of $g_i$. 

Fig.~\ref{fig:sig_bg_bar} tells us that the ILC with unpolarized beam
can reach small LFV couplings for $\sqrt{s}=1~\mathrm{TeV}$ compared to
the limits from the Belle experiment.
Even for lower energy such as $\sqrt{s}=250~\mathrm{GeV}$, as planned
as a first stage of the ILC experiment, 
requiring the initial $e^-$ beam to be right-handed makes the
experiment sensitive to smaller LFV couplings for $g_2, g_3, g_5$ and 
$g_6$ in the effective Lagrangian~(\ref{eq:efflag}).
The sensitivity of the ILC for the LFV contact interactions might be
competitive with 
the LFV search at the super-KEKB experiment.
The expected upper limits on the branching fractions of the $\tau$ LFV
decays at the Belle II can be found in \cite{Kou:2018nap}, which are
extrapolated from Belle results assuming 
$L_\mathrm{int}=50~\mathrm{ab}^{-1}$. 
The upper limit of $\mathrm{Br}(\tau \to 3e)$ in (\ref{belle_eee}) 
at the Belle experiment, which has been obtained using 
$L_\mathrm{int}=782~\mathrm{fb}^{-1}$~\cite{Hayasaka:2010np}, 
will be improved to be $\mathrm{Br}(\tau \to 3e) \sim 4.2 \times
10^{-8}$.
Then, the limits on the LFV couplings $g_i$ are 
$\sqrt{4.2\times 10^{-10}/2.7\times 10^{-8}}
\sim 0.12$ times smaller than those from the Belle experiment.
We give the expected upper limits of $g_i$ at the Belle II experiment
in Table~\ref{table:limits}.
We find that, for $\sqrt{s}=1~\mathrm{TeV}$ and $P(e^-)=0.8$, the upper
limits of $g_5$ at the ILC is slightly better than that of Belle II
while $g_2$ and $g_3$ are competitive.
If the $e^+$ beam is polarized in addition to $e^-$, i.e.,
$P(e^-)=0.8$ and $P(e^+)=-0.3$, the limits on $g_2,g_3$ and $g_5$ at the
ILC with $\sqrt{s}=1~\mathrm{TeV}$ are competitive with the Belle II 
at $50~\mathrm{ab}^{-1}$. 

\begin{figure}[t]
  \begin{center}
   \includegraphics[clip, width=8cm]{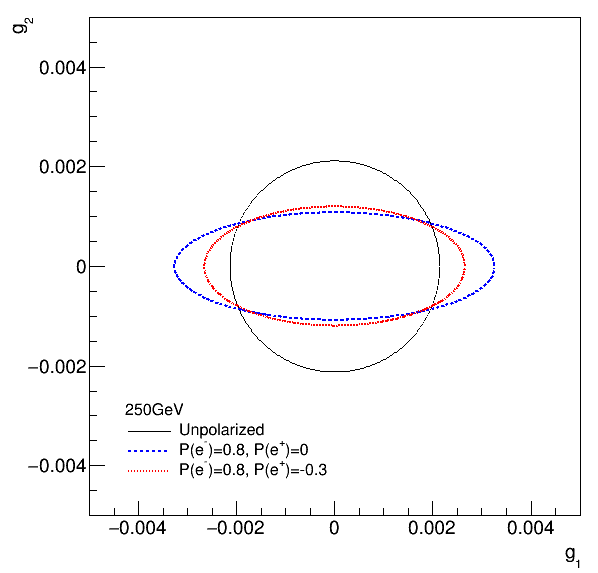}
   \caption{
   Constraints on the scalar LFV couplings $g_1$ and $g_2$ for
   $\sqrt{s}=250~\mathrm{GeV}$. The solid, dotted and dashed lines
   correspond to the unpolarized beam, $P(e^-)=0.8,~P(e^+)=0$, and 
   $P(e^-)=0.8,~P(e^+)=-0.3$, respectively.    
   }
    \label{fig:pol250gev}
  \end{center}
\end{figure}

We briefly discuss the advantage to use the polarization beam for the
LFV search at the ILC.
Throughout our analysis,  
we have been based on so called the single-coupling dominant
hypothesis, in which only one of six LFV couplings is finite and the 
rest are assumed to be zero.
However, to be realistic, it is necessary to study the case where
multiple LFV couplings exist.
For example, let us assume that a scalar particle mediates the LFV 
processes so that $g_1,~g_2 \neq 0$ and $g_i=0~(i=3\sim 6)$ in the
effective Lagrangian~(\ref{eq:efflag}).
We compare in Fig.~\ref{fig:pol250gev} the upper limits at 95\% CL 
on $(g_1,~g_2)$ plane from (i) unpolarized beam, (ii)
$P(e^-)=0.8,~P(e^+)=0$, and (iii) $P(e^-)=0.8,~P(e^+)=-0.3$ for
$\sqrt{s}=250~\mathrm{GeV}$.
As shown in the figure, the unpolarized beam constrains the combination 
$g_1^2 + g_2^2$. The polarized beam, however, constrains the LFV
coupling $g_2$ much severe than the unpolarized case.
This demonstrates that the beam polarization at the ILC can achieve 
different sensitivities to $g_1$ and $g_2$ by a cross-section
measurement.

We have so far discussed constraints on the LFV processes at the ILC
without considering specific new physics models. 
It is easy to read our constraints on the LFV couplings as those on new
physics models with scalar or vector mediator.
Suppose the following interactions of a flavored scalar $S$ or $Z'$: 
\begin{align}
 \mathcal{L} &\supset 
 y^\alpha_{\ell } \overline{\ell}P_\alpha e S,~
 q^\alpha_{\ell} g_{Z'} \overline{\ell} \gamma^\mu P_\alpha e Z'_\mu,
 \label{eq:model}
\end{align}
where $\ell$ represents $e$ or $\tau$, and $\alpha=L,~R$ denotes the 
chirality of the electron.
The gauge coupling of $Z'$ boson is denoted by $g_{Z'}$, and
$y^\alpha_\ell$ and $q^\alpha_\ell$ are dimensionless couplings.
In terms of couplings in eq.~(\ref{eq:model}), 
the LFV couplings in the effective Lagrangian (\ref{eq:efflag}) can be
expressed as
  \begin{align}
   g_i &\approx 0.031 \times \frac{y^\alpha_\tau y^\beta_e}
   {\left(m_S/1~\mathrm{TeV}\right)^2}~~(i=1,2)
   \\
   &\approx 0.031 \times g_{Z'}^2
   \frac{q_\tau^\alpha q_e^\beta}
   {\left(m_{Z'}/1~\mathrm{TeV}\right)^2}~~(i=3\sim 6), 
  \end{align}
  where $m_S$ and $m_{Z'}$ denote the scalar and $Z'$ boson mass,
  respectively.
Our results are valid for a mediator mass $M$ much larger than the
  momentum transfer $q$, i.e. $q^2 \ll M^2$.
We adopted this
  condition because no signature of new physics has been found at the
  LHC yet. However, if the mediator particle is baryophobic, it could
  not be
  produced at the LHC and its mass can be as small as the momentum
  transfer at the ILC. Then properties of the mediator particle could be
  studied at the ILC using, for example, energy and/or angular
  distributions of the final states. 
  Such a scenario might be another possibility of the LFV processes at
  the ILC but beyond the scope of our paper. 
  

 \section{Summary}
We have studied the possibility of the ILC to search for the LFV process
$e^+ e^- \to e^+ \tau^-$ via the four-Fermi contact interactions. 
Taking account of event selection conditions and polarization effects of
the initial $e^-$ beam, we found that the ILC could improve 
the upper bounds on the LFV couplings 
from the the measurement of $\mathrm{Br}(\tau \to 3e)$ by the Belle
collaboration.
In these couplings,
several factors of improvement is expected for 
$\sqrt{s}=250~\mathrm{GeV}$, and the improvement becomes an order of 
magnitude for $\sqrt{s}=1~\mathrm{TeV}$ with beam polarization. 
It is important to discriminate the dominant LFV couplings $g_1\sim g_6$ 
once the LFV process is found in some experiments. This can be partly
achieved by investigating the signal strength with different beam
polarization as we demonstrated for the scalar couplings $g_1$ and $g_2$. 
We note that there are other proposals for discrimination on the LFV
contact interactions using the polarization of $\tau$-lepton in the
$\tau^+ \tau^-$ production process~\cite{Kitano:2000fg,Goto:2010sn}.

\section*{Acknowledgments}
G.C.C is grateful to K. Hayasaka, Y. Okumura and Y. Yamamoto for
discussions and comments. 
The work of G.C.C is supported in part by Grants-in-Aid for Scientific 
Research from the Japan Society for the Promotion of Science (No.16K05314).  

 
%
\end{document}